\documentclass[twocolumn,letterpaper]{IEEEAerospaceCLS}
\usepackage{amsmath,amssymb,amsfonts}
\usepackage{algorithmic}
\usepackage{graphicx}
\usepackage{textcomp}
\usepackage{xcolor}
\usepackage{booktabs}
\usepackage{siunitx}
\usepackage{array}
\usepackage{tabularx}
\usepackage{capt-of}
\usepackage{tikz}
\usepackage{url}
\usepackage{cite}
\makeatletter
\renewcommand{\fnum@figure}{Fig.~\thefigure}
\makeatother
\usetikzlibrary{positioning,arrows.meta,shapes.multipart}
\DeclareSIUnit{\foot}{ft}
\usetikzlibrary{angles,quotes,arrows.meta,calc}
\newcommand{\N}{\ensuremath{\mathcal{N}}}
\newcommand{\U}{\ensuremath{\mathcal{U}}}

\newcommand{\trans}{\mathsf{T}}
\DeclareMathOperator{\clip}{clip}
\newcommand{\tbd}[1]{{\color{red}\ifx\relax#1\relax\textbf{TBD}\else#1\fi}}

\def\BibTeX{{\rm B\kern-.05em{\sc i\kern-.025em b}\kern-.08em
    T\kern-.1667em\lower.7ex\hbox{E}\kern-.125emX}}
\begin{document}
\renewcommand{\topfraction}{0.9}
\renewcommand{\dbltopfraction}{0.9}
\setcounter{dbltopnumber}{2}
\renewcommand{\dblfloatpagefraction}{0.8}

\title{Azimuth Pointing and Momentum Management for a Lightweight High-Altitude Balloon Payload}

\author{%
Yorgo Abou Haidar\\
Dept. of Mechanical Engineering\\
McGill University\\
817 Sherbrooke Street West\\
Montreal, QC, Canada H3A 0C3\\
yorgo.abouhaidar@mail.mcgill.ca
\and
James Richard Forbes\\
Dept. of Mechanical Engineering\\
McGill University\\
817 Sherbrooke Street West\\
Montreal, QC, Canada H3A 0C3\\
james.richard.forbes@mcgill.ca
\and
Andrew Higgins\\
Dept. of Mechanical Engineering\\
McGill University\\
817 Sherbrooke Street West\\
Montreal, QC, Canada H3A 0C3\\
andrew.higgins@mcgill.ca
\and
Benedetto D'Agnillo\\
Dept. of Mechanical Engineering\\
McGill University\\
817 Sherbrooke Street West\\
Montreal, QC, Canada H3A 0C3\\
benedetto.dagnillo@mail.mcgill.ca
\and
Justin Albert\\
Dept. of Physics \& Astronomy\\
University of Victoria\\
3800 Finnerty Road\\
Victoria, BC, Canada V8P 5C2\\
jalbert@uvic.ca
\thanks{\footnotesize \copyright~2027 IEEE. Personal use of this material is permitted. Permission from IEEE must be obtained for all other uses, in any current or future media, including reprinting/republishing this material for advertising or promotional purposes, creating new collective works, for resale or redistribution to servers or lists, or reuse of any copyrighted component of this work in other works.}
}
\maketitle
\thispagestyle{plain}
\pagestyle{plain}
\begin{abstract}
The ALTAIR (Airborne Laser for Telescopic Atmospheric Interference Reduction) project provides photometric calibration for ground-based telescopes by flying an integrating sphere as an artificial star-like calibration source on a high-altitude balloon. Because the angular distribution of light from the exit port of the sphere is approximately Lambertian, the payload must keep its azimuth within $\pm8^\circ$ of the telescope line of sight. This condition bounds the fractional difference in light intensity from that along the exit-port normal to below \SI{1}{\percent}. The bound must be held for at least 30~s at a time. This paper presents a pointing architecture that combines passive stabilization, onboard attitude determination, and active azimuth control using a momentum wheel, with active desaturation through a second motor at the flight-train pivot. Carbon-fiber inertial booms reduce off-axis pendulating motion sufficiently for azimuth control to be treated independently. Attitude determination is performed using onboard sensor fusion of rate gyros, accelerometers, magnetometers, and carrier-phase measurements from dual Global Navigation Satellite System (GNSS) antennas. A dynamic model of the payload is simulated in a high-fidelity environment including motor dynamics and limits, sensor noise, and wind disturbances generated from flight data. The simulation is used to size actuators and tune the control loops. The payload avionics, flight software, power management, controls, and real-time ground telemetry systems are developed in-house. The system is evaluated through ground testing, simulation, and two flights. With the momentum wheel alone, 53 intervals of 30~s pointing within $\pm8^\circ$ of the target azimuth are recorded during ascent, with a root-mean-square error (RMSE) of $3.7^\circ$, and the longest uninterrupted window lasts 7~min 55~s with $5^\circ$ of pointing RMSE. Simulation results suggest significant performance improvements from adding a pivot motor for momentum management.
\end{abstract}

\tableofcontents

\section{Introduction}
The ALTAIR (Airborne Laser for Telescopic Atmospheric Interference Reduction) project is a collaboration between the McGill University Mechanical Engineering Department and the University of Victoria Physics Department. The objective is to provide precise photometric calibration for ground-based telescopes by flying calibrated, real-time, \textit{in situ} optically measured light sources on high-altitude balloons. Precise flux calibration is a major limiting factor in precision cosmology: percent-level uncertainties in photometric zero points translate directly into large uncertainties in dark-energy equation-of-state parameters \cite{Stubbs2006Toward1Percent}. ALTAIR addresses this challenge using an integrating sphere as a well-characterized, approximately Lambertian optical source, elevated to between 15 and 30 \si{\kilo\metre} to act as a monochromatic artificial star-like calibration source at a variety of wavelengths with SI-traceable photometric output.

\begin{figure}[htb]
\centering
\includegraphics[width=\columnwidth]{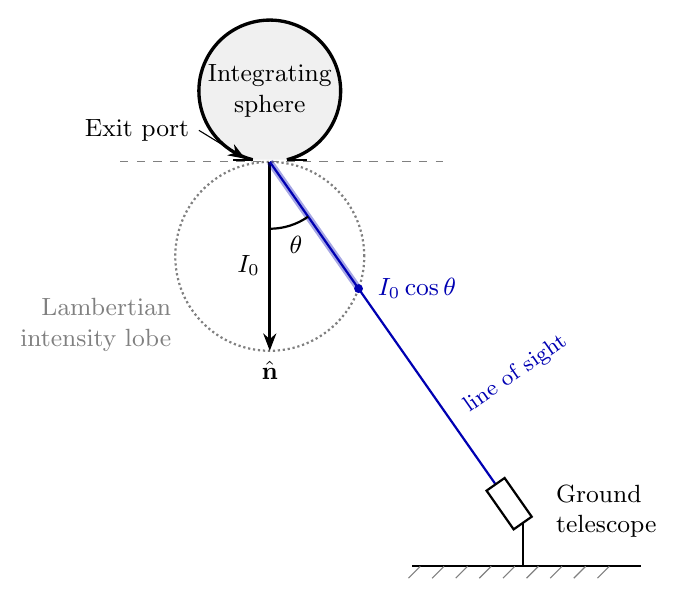}
\caption{Lambertian emission from the integrating-sphere exit port. The radiant intensity toward the telescope falls off as $I_0\cos\theta$, where $I_0$ is the radiant intensity along the exit-port normal $\hat{\mathbf{n}}$ and $\theta$ is the angle between $\hat{\mathbf{n}}$ and the telescope line of sight.}  \label{fig:lambertian}
\end{figure}

The exit port of the light source is approximately Lambertian, so the light flux measured by the telescope is proportional to the cosine of the angle $\theta$ between the exit-port direction and the telescope line of sight, as illustrated in Fig.~\ref{fig:lambertian}. 
Rather than reconstructing $\theta(t)$ and integrating the cosine correction over a multi-second exposure image, the flux is reduced under the assumption of perfect pointing, and the pointing requirement is set such that the error incurred by that assumption is bounded. Since the flux deficit is bounded at every instant,
\begin{equation}
  1-\cos\theta \le 1-\cos\theta_{\mathrm{max}},
  \label{eq:flux_bound}
\end{equation}
setting $\theta_{\mathrm{max}} = \ang{8}$ bounds this deficit to \SI{0.97}{\percent} of the intensity along the exit-port normal, regardless of how $\theta$ varies within the exposure. Laboratory calibration further shows that the angular emission profile of the sphere is best characterized as Lambertian near the exit's normal direction as defined in Fig.~\ref{fig:lambertian}, so pointing keeps the source in the region where its output is most reliably known. Pointing duration, defined as the longest uninterrupted interval over which $\theta < \theta_{\mathrm{max}}$ is maintained, is an equally important performance metric: ALTAIR requires at least \SI{30}{\second} of continuous pointing so that multi-second exposure images can be taken. 

Precise azimuth control of large balloon-borne payloads has been extensively studied \cite{Benford2012PrecisionAttitudeControlBETTII, Didier2015EBEXPointing, Kassarian2024FIREBallRobustLOS}. A payload developed at the University of California Santa Barbara by Chingcuanco \emph{et~al.}~achieved \ang{0.02} RMS error with a momentum wheel pointing system and a momentum management capability~\cite{Chingcuanco1990ModelingControlBalloonPlatform}. The Wallops Arc Second Pointer and the Balloon-borne Imaging Testbed are two examples of large payloads achieving sub-arcsecond azimuth pointing RMS error~\cite{stuchlik2015_wasp, romualdez2016_bit_pointing}. These gondolas have masses of hundreds of kilograms, with correspondingly large inertia.

Canadian aviation regulations state that a balloon payload is considered \emph{medium} in size and does not require authorization from the Minister of Transport for flight if the balloon volume is under \SI{115}{\cubic\foot} \cite{canadian_aviation_regulations_1996}. This volume corresponds to \SI{3.26}{\cubic\metre} and provides roughly \SI{3.6}{\kilo\gram} of lift at sea level. For comparison, the payloads discussed previously were flown using balloons with volumes of \SI{e5}{\cubic\metre} and \SI{e6}{\cubic\metre}~\cite{Chingcuanco1990ModelingControlBalloonPlatform, romualdez2016_bit_pointing}. The BUBBLE (Buoyancy Balloon Bus Lifted Experiments) payload is a baseline example of flywheel-based azimuth control aboard a lightweight payload. The experiment reports errors of \ang{15}\textendash\ang{20} with a maximum continuous pointing window of \SI{140}{\second}~\cite{Zietz2024BUBBLE_RWOnly}. Wheel saturation was the limiting factor preventing the BUBBLE experiment from reaching longer pointing windows. 

This paper will present the design of a flight-proven azimuth pointing architecture for a ``medium'' class payload, as defined by Canadian regulations, in order to meet the ALTAIR experiment's pointing requirements. Focus is placed on simulation development, passive stabilization and the design of a momentum management system. Section~\ref{sec:architecture} will outline the payload's design, its pointing system, and attitude determination sensors. Sections~\ref{sec:modeling} and~\ref{sec:control} will focus on the payload's attitude dynamics, actuator and sensor models, wind disturbance modeling and controller design that were simulated in a Python environment. Section~\ref{sec:results} will present and discuss results from simulation and flight tests. Section~\ref{sec:lessons} summarizes the contributions of this work and outlines future work.

\section{System Architecture}
\label{sec:architecture}

\begin{figure*}[t]
\centering
\includegraphics[width=\textwidth]{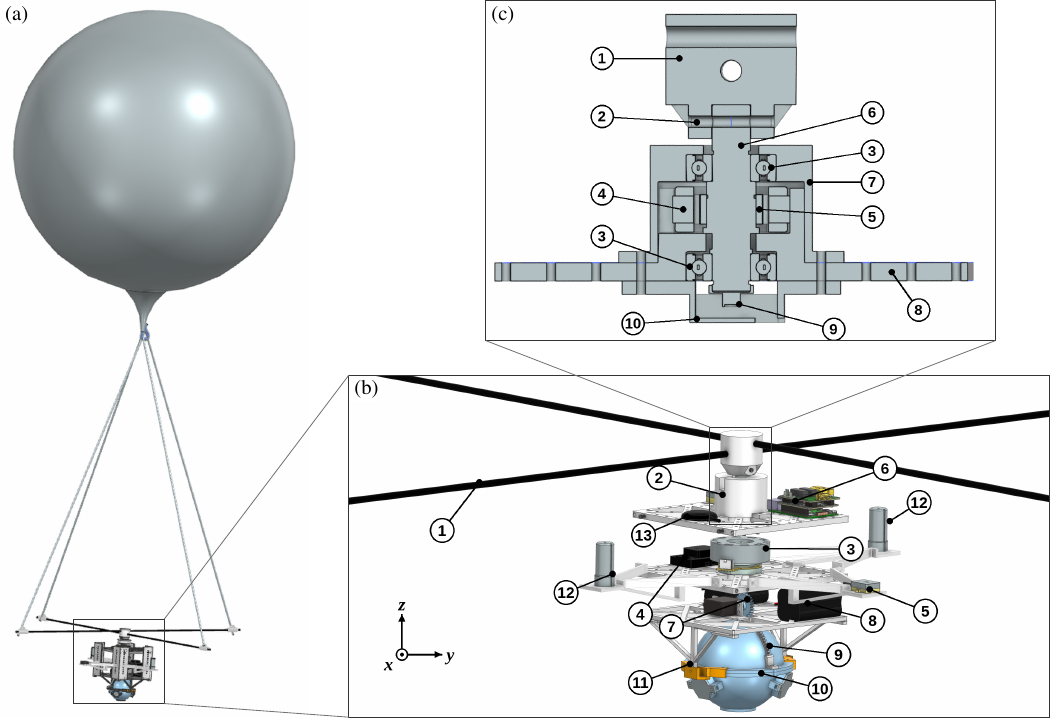}\\[6pt]
\begin{tabular*}{\textwidth}{@{\extracolsep{\fill}}llll@{}}
\multicolumn{2}{@{}l}{\textit{(b) Payload}} & \multicolumn{2}{l@{}}{\textit{(c) Pivot Motor}} \\
(1)~carbon-fiber rods           & (8)~batteries            & (1)~rod connector & (6)~shaft \\
(2)~pivot motor                 & (9)~sphere drive chain   & (2)~screw hole    & (7)~motor housing \\
(3)~momentum wheel              & (10)~integrating sphere  & (3)~ball bearings & (8)~payload top deck \\
(4)~Pixhawk flight controller   & (11)~sphere mounting rig & (4)~motor stator  & (9)~magnet mount \\
(5)~external magnetometer       & (12)~GNSS antennas       & (5)~motor rotor   & (10)~encoder mount \\
(6)~flight computer             & (13)~secondary GNSS      &                   & \\
(7)~servo gear motor            &                          &                   & \\
\end{tabular*}
\caption{ALTAIR flight configuration: (a) balloon, rigging, and payload, (b) payload CAD model with side panels hidden, and (c) cross-section of the motorized pivot.}
\label{fig:system}
\end{figure*}

An annotated computer-aided design (CAD) of the ALTAIR payload can be seen in Fig.~\ref{fig:system}(b). It consists of multiple subsystems developed by the ALTAIR team including custom avionics, a telemetry system (antenna not included in this CAD drawing) and a modular payload structure. These systems enable the experiment to run autonomously: the telemetry system uses LoRa~\cite{semtech_lora} to relay its coordinates to a ground station, which points a telescope toward the balloon, tracking and imaging it without operator input. 
The payload is passively stabilized using carbon-fiber rods that serve a double purpose: increasing the payload inertia and increasing inertial coupling of the payload and the balloon by rigging each rod's extremity to the balloon neck directly as demonstrated by Fig.~\ref{fig:system}(a). 

\subsection{Actuators}

The payload includes three actuators that each serve a specific purpose in pointing the integrating sphere. A servo with an angular resolution of \ang{1} pitches the sphere via a $4{:}1$ gear reduction, giving a final angular resolution of \ang{0.25}. A brushless DC motor with a mounted custom flywheel design serves as a momentum wheel, accelerating and decelerating to torque the payload, controlling its azimuth angle. As the momentum wheel speeds up or down it reaches a saturated state where it may not accelerate past its maximum speed or decelerate and cross zero speed. This constraint means limited control authority for the motor and causes interruptions in pointing. To address this problem, a pivot motor is built at the connection between the payload and flight train to torque the payload against the balloon and flight train system, allowing the momentum wheel to gradually dump its stored angular momentum and maintain pointing authority~\cite{Chingcuanco1990ModelingControlBalloonPlatform, shariff2014spider}. A brushless DC motor is used for the pivot motor. The housing and shaft are designed such that the joint can reliably support the weight of the entire payload while allowing decoupling of the payload and flight train. A cross-sectional view of the pivot motor's components is displayed in Fig.~\ref{fig:system}(c).

\subsection{Attitude Determination Hardware} 

Accurate determination of the payload azimuth angle in a known reference frame is required both to direct the sphere exit port toward a ground telescope and to reconstruct the attitude history for post-flight analysis. A Pixhawk 6X flight controller provides the core of the attitude determination system, selected for its ecosystem of supported sensors and integrated sensor fusion. The device has three built-in 6-axis inertial measurement units (IMUs) and a magnetometer that it uses to estimate its three-dimensional attitude in a north-east-down (NED) frame. Estimating the payload's heading relative to north relies on the magnetometer that measures the earth's magnetic field to determine where magnetic north lies, providing an inertial heading reference. Due to magnetic interference from the motors, external sensors are needed to complement the built-in magnetometer. An external magnetometer was therefore mounted on an arm to distance it from the sources of interference, as shown in Fig.~\ref{fig:system}(b). 

The added magnetometer was still vulnerable to interference and so another heading solution was pursued. Using two GNSS antennas separated by a known distance and whose baseline is defined in the payload body frame, the difference in carrier-phase of a GNSS signal can be used to estimate payload heading. These antennas can also be seen annotated in Fig.~\ref{fig:system}(b), they are extended outside the payload on arms to provide a longer baseline for more precise estimation.

\section{Physics Modeling and Simulation}
\label{sec:modeling}
The physics models developed were simulated in Python and used to size actuators, test guidance algorithms, tune controllers, and predict flight performance. Three assumptions are made to simplify the modeling process:
\begin{enumerate}
    \item The center of mass cannot be perfectly centered, and this creates coupling between the payload azimuth and pendulating motion. When the payload is tilted, the flight train's tension no longer passes through the center of mass, and this creates a persistent torque that the momentum wheel cannot offset, disrupting pointing. In the following derivations, it is assumed that the payload is stable about its $x$ and $y$ axes as defined in Fig.~\ref{fig:system}(b), so the payload azimuth dynamics are treated independently. Pendulating motion is quantified by calculating the RMS of the payload's pitch and roll angles and is referred to as \emph{tilt}. Tilt stability is resolved by adding inertial booms and tightly coupling them to the balloon neck. The stability assumption is further reinforced at float altitude, where wind disturbances are substantially lesser than those during ascent. As a result, the models developed for simulation have been reduced to rotation about a single axis. The tilt stability of the payload is measured in flight by the onboard IMUs and is key in quantifying flight performance. 
    \item The payload and flight-train rotations are decoupled. The rotations are only related through the pivot motor's Coulomb friction term.
    \item The balloon is approximated to a momentum \emph{sink} due to its large inertia and surface-area damping. As a result, its dynamics are omitted. This assumption is made in previous work \cite{Chingcuanco1990ModelingControlBalloonPlatform} for a similar purpose.
\end{enumerate}

\subsection{Attitude Dynamics}
The generalized coordinate vector is written as 
\begin{equation}
  \label{eq:coords}
  q = \begin{bmatrix} \alpha & \beta & \gamma \end{bmatrix}^{\trans},
\end{equation}
where $\alpha$ is the azimuth of the payload, $\beta$ that of the momentum wheel, and $\gamma$ that of the pivot motor shaft rigidly connected to the flight train, all resolved in the same inertial frame and expressed in \si{\radian}. The flight train does not have a direction as such, instead $\gamma=0$ is defined as the flight train's unwound state.

The dynamic equation for the payload is written as 
\begin{equation}
    \label{eq:payload_eom}
    J_{\mathrm{p}}\ddot{\alpha} = -\tau_{\mathrm{mw}} - \tau_{\mathrm{pm}} + \tau_{\mathrm{d}} - C_{\mathrm{p}}\dot{\alpha},
\end{equation}
where $J_{\mathrm{p}}$ is the payload inertia about the axis of rotation (\textit{z}-axis) in \si{\kilo\gram\metre\squared}, $\tau_{\mathrm{mw}}$ is the torque applied by the momentum wheel, $\tau_{\mathrm{pm}}$ is the torque applied by the pivot motor, $\tau_{\mathrm{d}}$ is the torque applied by wind disturbances, where all torques are in \si{\newton\metre}. The payload damping coefficient is $C_{\mathrm{p}}$, in \si{\newton\metre\second\per\radian}.

The momentum wheel obeys
\begin{equation}
    \label{eq:flywheel_eom}
    J_{\mathrm{mw}}\ddot{\beta} = \tau_{\mathrm{mw}} - C_{\mathrm{mw}}\dot{\beta},
\end{equation}
where $J_{\mathrm{mw}}$ is the momentum wheel inertia in \si{\kilo\gram\metre\squared} and $C_{\mathrm{mw}}$ is the momentum wheel damping coefficient in \si{\newton\metre\second\per\radian}.

The pivot motor shaft obeys
\begin{equation}
    \label{eq:pivot_eom}
    J_{\mathrm{ft}}\ddot{\gamma} = \tau_{\mathrm{pm}} - K_{\mathrm{ft}}\gamma - C_{\mathrm{ft}}\dot{\gamma},
\end{equation}
where $J_{\mathrm{ft}}$ is the flight-train (including the inertial booms and pivot motor shaft) moment of inertia in \si{\kilo\gram\metre\squared}, $K_{\mathrm{ft}}$ is the stiffness of the rigging in \si{\newton\metre\per\radian}, and $C_{\mathrm{ft}}$ is the flight-train damping coefficient in \si{\newton\metre\second\per\radian}.

\subsection{Motor Modeling}
The transient brushless DC motor models are described by the governing equations
\begin{align}
    \tau_{\mathrm{motor}} &= K_{\mathrm{t}} I - b\omega - f_{\mathrm{C}}\operatorname{sgn}(\omega), \label{eq:mechanical_ode}\\
    L\frac{\mathrm{d}I}{\mathrm{d}t} &= V - RI - K_{\mathrm{b}}\omega, \label{eq:electrical_ode}
\end{align}
where $V$, $I$, $R$, and $L$ are the motor voltage, current, resistance, and inductance respectively in \si{\volt}, \si{\ampere}, \si{\ohm} and \si{\henry}, $\tau_{\mathrm{motor}}$ is the motor torque output in \si{\newton\metre}, $K_{\mathrm{t}}$ is the torque constant in \si{\newton\metre\per\ampere}, $K_{\mathrm{b}}$ is the back-electromotive force (EMF) constant in \si{\volt\second\per\radian}, $\omega$ is the rotor speed relative to the stator in \si{\radian\per\second}, $b$ is the viscous damping coefficient in \si{\newton\metre\second\per\radian}, and $f_{\mathrm{C}}$ is the Coulomb friction in \si{\newton\metre}.

The effect of motor back-EMF saturation, the main constraint on momentum wheel control, can be inferred from the equations. Since the electrical time constant $L/R$ is much smaller than the mechanical time constant, the current settles almost instantly relative to changes in rotor speed. The inductive term $L\,\mathrm{d}I/\mathrm{d}t$ is therefore neglected, and \eqref{eq:electrical_ode} reduces to $I = (V - K_{\mathrm{b}}\omega)/R$. Substituting $I$ into \eqref{eq:mechanical_ode}, the motor torque output becomes 
\begin{equation}
    \tau_{\mathrm{motor}}(\omega) =  \frac{K_{\mathrm{t}}}{R}V - \frac{K_{\mathrm{t}}K_{\mathrm{b}}}{R}\omega, \label{eq:saturation}
\end{equation}
where $\tau_{\mathrm{motor}}$ is the torque applied to the flywheel. For a given $V=V_{\mathrm{max}}$ derived from mechanical and power constraints, the angular velocity of the flywheel increases until $(K_{\mathrm{t}}/R)V_{\mathrm{max}}=(K_{\mathrm{t}}K_{\mathrm{b}}/R)\,\omega$, at which point the torque output becomes zero. Once the motor is unable to torque in a given direction, it loses control authority and must dump momentum.

\subsection{Sensor Modeling}
Modeling the sensors aboard the payload, including their noise properties and sampling rates, is useful for improving the simulation fidelity and determining controller robustness. The gyroscope noise is characterized from \SI{8}{\hour} of static data recorded at \SI{50}{\hertz}, from which an Allan deviation analysis \cite{fusionportable_allan_variance_ros} yields an angular random walk $\sigma_{\mathrm{v}}$ in \si{\degree\per\sqrt{\second}} and a rate random walk $\sigma_{\mathrm{u}}$ in \si{\degree\per\second\per\sqrt{\second}}. The simulated gyroscope measurement $\tilde{\omega}$ is corrupted by white noise and a random-walk bias,
\begin{equation}
    \tilde{\omega}_k = \omega_k + b_k + \eta_{\mathrm{v},k}, \qquad
    b_{k+1} = b_k + \eta_{\mathrm{u},k},
    \label{eq:gyro_model}
\end{equation}
where $\eta_{\mathrm{v},k} \sim \mathcal{N}\!\left(0,\, \sigma_\mathrm{v}^2/\Delta t\right)$ is the white measurement noise, $b_k$ is a random-walk bias driven by the white noise $\eta_{\mathrm{u},k} \sim \mathcal{N}\!\left(0,\, \sigma_\mathrm{u}^2\,\Delta t\right)$, and $\Delta t$ is the sampling period.

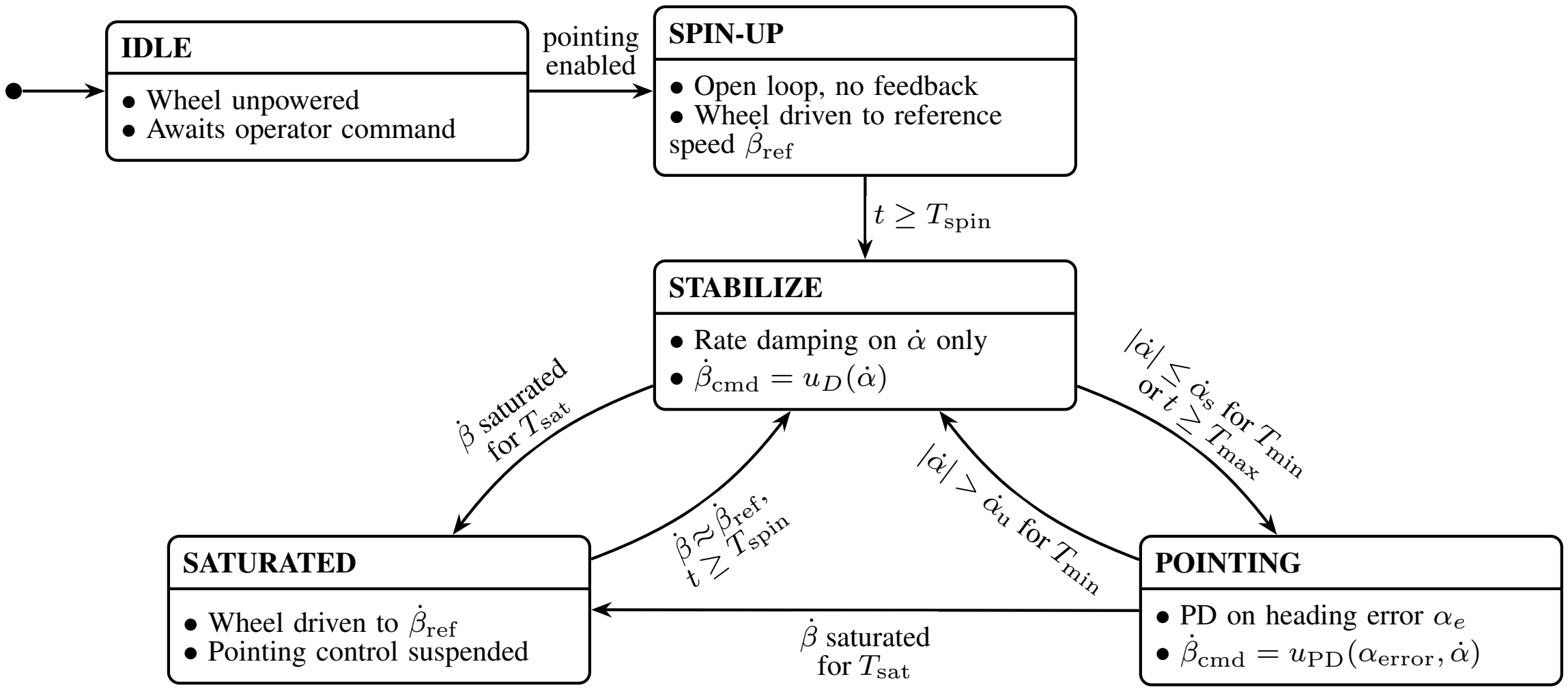
\begin{figure*}[!t]
\centering
\begin{tikzpicture}[
  font=\footnotesize,
  node distance=8mm,
  state/.style={draw, thick, rounded corners=3pt,
                rectangle split, rectangle split parts=2,
                rectangle split part align=left,
                text width=38mm, align=left, inner sep=4pt},
  trans/.style={-{Stealth[length=2mm]}, thick},
  lbl/.style={font=\scriptsize, align=center, inner sep=2pt, fill=white}
]

\node[state] (spin)
  {\textbf{SPIN-UP}
   \nodepart{two}$\bullet$~Open loop, no feedback\\
                 $\bullet$~Wheel driven to reference speed $\dot\beta_{\mathrm{ref}}$};

\node[state, left=12mm of spin] (idle)
  {\textbf{IDLE}
   \nodepart{two}$\bullet$~Wheel unpowered\\
                 $\bullet$~Awaits operator command};

\node[state, below=8mm of spin] (stab)
  {\textbf{STABILIZE}
   \nodepart{two}$\bullet$~Rate damping on $\dot\alpha$ only\\
                 $\bullet$~$\dot\beta_{\mathrm{cmd}}=u_D(\dot\alpha)$};

\node[state, below left=12mm and 6mm of stab] (sat)
  {\textbf{SATURATED}
   \nodepart{two}$\bullet$~Wheel driven to $\dot\beta_{\mathrm{ref}}$\\
                 $\bullet$~Pointing control suspended};

\node[state, below right=12mm and 6mm of stab] (point)
  {\textbf{POINTING}
   \nodepart{two}$\bullet$~PD on heading error $\alpha_e$\\
                 $\bullet$~$\dot\beta_{\mathrm{cmd}}=u_{\mathrm{PD}}(\alpha_\mathrm{error},\dot\alpha)$};

\node[circle, fill=black, inner sep=1.6pt, left=8mm of idle] (init) {};

\draw[trans] (init) -- (idle);
\draw[trans] (idle) -- node[lbl,above=2pt]{pointing \\ enabled} (spin);
\draw[trans] (spin) -- node[lbl,right]{$t\ge T_{\mathrm{spin}}$} (stab);

\draw[trans] (stab) to[bend left=16] node[lbl,sloped,above,pos=0.5]
  {$|\dot\alpha|\le\dot\alpha_{\mathrm{s}}$ for $T_{\min}$\\ or $t\ge T_{\max}$} (point);
\draw[trans] (point) to[bend left=16] node[lbl,sloped,below,pos=0.5]
  {$|\dot\alpha|>\dot\alpha_{\mathrm{u}}$ for $T_{\min}$} (stab);

\draw[trans] (stab) to[bend right=16] node[lbl,sloped,above,pos=0.5]
  {$\dot\beta$ saturated\\ for $T_{\mathrm{sat}}$} (sat);
\draw[trans] (sat) to[bend right=16] node[lbl,sloped,below,pos=0.5]
  {$\dot\beta\!\approx\!\dot\beta_{\mathrm{ref}}$,\\ $t\ge T_{\mathrm{spin}}$} (stab);

\draw[trans] (point.west) -- node[lbl,below]
  {$\dot\beta$ saturated\\ for $T_{\mathrm{sat}}$} (sat.east);

\end{tikzpicture}
\caption{Azimuth pointing state machine for momentum-wheel-only control. Here, $\dot\beta$ is the wheel speed, $\dot\beta_{\mathrm{ref}}$ the reference wheel speed, and $\alpha$ the payload heading. The $\dot\alpha$  is the median-filtered angular rate over a \SI{0.5}{\second} window. Saturation is declared when $\dot\beta$ leaves the admissible band about $\dot\beta_{\mathrm{ref}}$ for longer than $T_{\mathrm{sat}}$, at either the upper or the lower bound. Transition guards on $\dot\alpha$ use a hysteresis pair, $\dot\alpha_{\mathrm{s}} < \dot\alpha_{\mathrm{u}}$, each held for $T_{\min}$. This configuration prevents repeated switching between stabilization and pointing when the angular rate sits near the threshold. $t$ is the time since entering the mode.}
\label{fig:pointing_fsm}
\end{figure*}

\subsection{Wind Disturbance Modeling}
A model is developed and trained with real flight data to simulate disturbance torques on the payload. Accurately modeling various realistic disturbance patterns that reproduce flight-like results is crucial if the simulation is to be relied on for controller testing and tuning. 

The wind acts on the payload as an azimuth disturbance torque with two parts,
\begin{equation}
\tau_{\mathrm{d}}(t) = \tau_{\mathrm{s}}(t) + \tilde{\tau}(t),
\label{eq:wind_total}
\end{equation}
a steady torque $\tau_{\mathrm{s}}$ of constant sign and a zero-mean gust torque $\tilde{\tau}$. Both depend on the altitude $h(t) = h_0 + v_\mathrm{a} t$ of the ascent modeled under a constant ascent rate assumption.

The angular rate of a freely rotating payload about its axis of rotation is recorded with a gyroscope during flight, and numerical differentiation of this rate yields the angular acceleration $\ddot{\alpha}_{\mathrm{d}}$. Because the payload is uncontrolled during the experiment, $\tau_{\mathrm{d}} = J_{\mathrm{p}}\,\ddot{\alpha}_{\mathrm{d}}$ captures the net external torque about the axis of rotation.

The gust component is strongly non-Gaussian: once the altitude trend is removed, its excess kurtosis is 7.7. These heavy tails are the gusts that cause pointing excursions and drive the wheel toward saturation. Gust histories are therefore generated as iterative amplitude-adjusted Fourier transform surrogates~\cite{schreiber1996}, which preserve the power spectrum and amplitude distribution of the source while randomizing its Fourier phases, so each run receives a new time history.
The gust term is then restored as a function of altitude,
\begin{equation}
\tilde{\tau}(t) = \sigma_0\,\frac{\rho\big(h(t)\big)}{\rho_0}\,g(t),
\label{eq:gust_scaling}
\end{equation}
where $g(t)$ is the unit-variance surrogate, which sets the shape of the gusts, and $\sigma_0$ is the RMS gust torque at sea level, which sets their magnitude. The ratio $\rho/\rho_0$ reduces this magnitude with altitude, where $\rho$ is the atmospheric density and $\rho_0$ is its sea-level value. Wind speed fluctuation with altitude is neglected as a modeling choice.

The steady torque follows 
\begin{equation}
\tau_{\mathrm{s}}(t) = c_\rho\,\rho\big(h(t)\big),
\label{eq:steady_torque}
\end{equation}
where $c_\rho$ is a constant calculated from flight data. The total angular momentum of the payload about its axis of rotation is reconstructed, and its rate of change is taken over consecutive windows of 2-min duration. Over each window the gusts largely average out, so the rate of change of momentum gives the steady torque at that altitude, and dividing by the local density gives one estimate of $c_\rho$. The median of these estimates is used as the fitted value.  Unlike the gusts, however, it does not average to zero: its momentum accumulates in the wheel and drives it into saturation.

\section{Controller Design}
\label{sec:control}

The control architecture is split between two actuators, the motor that controls the momentum wheel and is responsible for precise pointing of the payload and the pivot motor that torques the payload off the flight train and is primarily responsible for desaturating the momentum wheel. This system is commonly found on large balloon payloads \cite{chingcuanco1991mrac, shariff2014spider, romualdez2016_bit_pointing}. These controllers both affect the payload's azimuth but act independently of one another. Each controller has a distinct control objective, and to ensure that one does not interfere with the other, the two controllers must be bandwidth separated. Otherwise, coupling of the controllers would cause oscillatory behavior~\cite{Chingcuanco1990ModelingControlBalloonPlatform}. Each controller acts on a different disturbance mode: high-frequency disturbances are rejected by the momentum wheel, while the low-frequency biases that drive the momentum wheel toward saturation are rejected by the pivot motor. Chingcuanco \emph{et~al.}~reached this conclusion from a root-locus analysis of the system's transfer functions~\cite{Chingcuanco1990ModelingControlBalloonPlatform}, and Romualdez \emph{et~al.}~showed through a Routh analysis of the coupled rotational dynamics that the system becomes unstable above a critical pivot motor gain~\cite{romualdez2016_bit_pointing}. In this work, bandwidth separation is achieved by bounding the pivot motor gains well below those of the momentum wheel controller.

\subsection{Momentum Wheel Controller}
The momentum wheel is controlled with a proportional-derivative controller, meaning the commanded wheel speed is a function of the azimuth error and its derivative. The controller is described by 
\begin{align}
    u_{\mathrm{PD}} &= k_\mathrm{p}(\alpha - \alpha_{\mathrm{ref}}) + k_\mathrm{d}\dot{\alpha} + \dot{\beta}_{\mathrm{ref}}, \\
    \dot{\beta}_{\mathrm{cmd}} &= \clip(u_{\mathrm{PD}}, 0, \dot\beta_{\mathrm{max}}),
\end{align}
where the controller output $\dot{\beta}_{\mathrm{cmd}}$ is the commanded momentum wheel speed, offset by a constant setpoint speed $\dot{\beta}_{\mathrm{ref}}$. This setpoint is taken as half the maximum motor speed, because the motor's control authority lies between zero and the maximum speed. The setpoint is nonzero to avoid zero-speed crossings, where stiction produces a torque discontinuity that disrupts pointing. When the control objective is rate stabilization instead of pointing, the proportional term is dropped and the derivative term is used alone with the same gain. The controller gains were tuned in simulation and during ground tests. 
This control strategy is different from traditional momentum wheel control, in which the motor torque is the controller output~\cite{Zietz2024BUBBLE_RWOnly,romualdez2016_bit_pointing}. The choice to control angular velocity directly is tailored to ALTAIR and assumes a number of conditions:
\begin{enumerate}
    \item The reference azimuth must be continuously and slowly varying. This controller cannot handle sudden jumps in desired pointing direction, such instances will create large initial discrepancies in commanded speed when the reference changes. This discrepancy will result in an undesired torque which will disrupt pointing.
    \item Once the pointing requirement is reached the controller cannot effectively reject small disturbance torques, instead the payload drifts from its set point slowly. When the momentum wheel is flown independently, the controller is tuned to ensure the drift is limited to an acceptable error and that the wheel quickly saturates if it cannot restore pointing. Otherwise, the pivot motor solves this issue by rejecting momentum build-up.
\end{enumerate}

\begin{figure*}[!t]
    \centering
    \includegraphics[width=\linewidth]{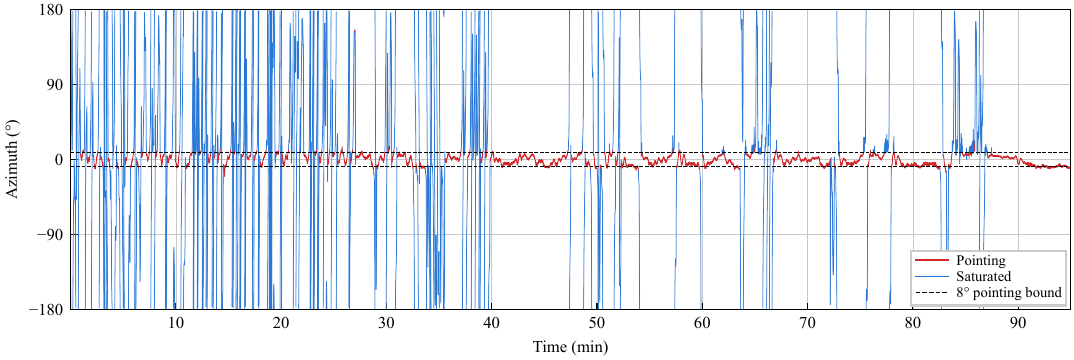}
    \caption{Azimuth time series for the ascent of Flight~2.}
    \label{fig:pointing}
\end{figure*}

\subsection{Desaturation Controller}
The pivot motor is controlled by a proportional controller that computes a current signal. The controller is described by
\begin{align}
    u_{\mathrm{P}} &= k_\mathrm{v}(\dot{\beta} - \dot{\beta}_{\mathrm{ref}}), \\
    I_{\mathrm{cmd}} &= \clip(u_{\mathrm{P}}, -I_{\mathrm{max}}, I_{\mathrm{max}}).
\end{align}
Desaturation requires only small pivot torques at low relative speeds. These torques lie near the minimum the motor can reliably produce, a limit set by stiction effects. The pivot motor is therefore actuated in short, infrequent pulses rather than continuously, and the actuation period and control loop rate become tuning parameters. The control loop period sets the magnitude of the commanded torque. A longer period allows the momentum wheel to accumulate more angular velocity between updates, so the pivot motor commands a larger restoring torque that exceeds the stiction limit. All tuning parameters were tuned in simulation and refined through ground tests.

\subsection{Control Modes}
It is still possible to accomplish azimuth control of the payload without the pivot motor. In the expected event of motor saturation, a maneuver is performed by the motor to desaturate itself and recover pointing authority. The state machine logic responsible for operating the momentum wheel independently includes three control modes: pointing, stabilize, and saturation. The logic behind each mode and switching between modes is described in Fig.~\ref{fig:pointing_fsm}.

When the full pointing system is flown with the pivot motor, only a pointing mode is needed with the described control laws. The expectation is that momentum management will allow the momentum wheel to operate continuously for an entire flight.

\section{Results and Discussion}
\label{sec:results}

\subsection{Simulation Results}
\begin{figure}[t]
\centering
\includegraphics[width=\columnwidth]{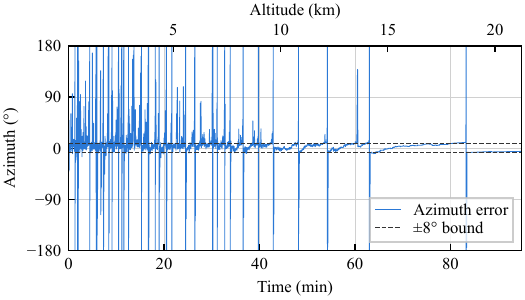}\\[-2pt]
{\footnotesize (a)}\\[4pt]
\includegraphics[width=\columnwidth]{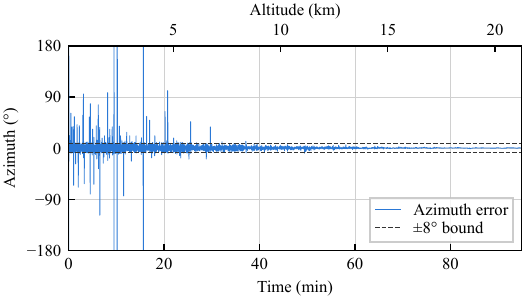}\\[-2pt]
{\footnotesize (b)}
\caption{Azimuth error for a simulated ascent of the payload: (a)~momentum wheel only (b)~momentum wheel with the pivot motor.}
\label{fig:sim_ascent}
\end{figure}

An entire flight ascent was simulated with only a momentum wheel controlled by the state machine described in Fig.~\ref{fig:pointing_fsm}. Flight results using the same pointing system are plotted in Fig.~\ref{fig:pointing} and help validate the simulation results plotted in Fig.~\ref{fig:sim_ascent}(a). The same saturation and recovery patterns observed in flight are reproduced in simulation. The disturbance pattern's altitude correlation is evident in both flight and simulation results. The azimuth drift observed in flight during long pointing windows is also correctly reproduced by the simulation, this behavior is expected and explained in Section~\ref{sec:control}. These results confirm that the modeled payload and disturbances are realistic and can be used to validate controllers for flight.

\begin{figure*}[t]
\centering
\includegraphics[width=\textwidth]{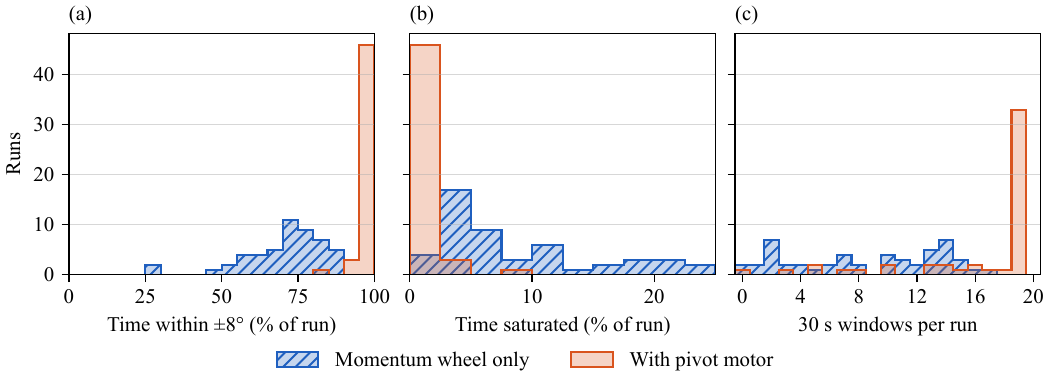}
\caption{Monte Carlo pointing performance over $n=50$ paired runs of 600~s, with and without the pivot motor: (a) percentage of the time with azimuth error within $\pm8^\circ$ (b) percentage of the time with the wheel saturated (c) number of 30~s intervals per run with the error continuously within $\pm8^\circ$.}
\label{fig:mc_windows}
\end{figure*}

The same ascent was simulated with the pivot motor active and without the state machine logic controlling the momentum wheel. Results are plotted in Fig.~\ref{fig:sim_ascent}(b). A 45 minute pointing window is achieved in simulation with the addition of the pivot motor, compared to 15 minutes with the momentum wheel only. These results confirm the benefit of pursuing a pivot motor for momentum management. 

Monte Carlo simulations are run to test controller robustness against uncertainties on simulated parameters. The disturbed parameters are listed in Table~\ref{tab:mc_dispersions}. The statistical distributions used to define the parameter space were determined using manufacturer specifications for the motors, and engineering judgment for the payload parameters. Each parameter is sampled independently for each of the 50 runs. 
\begin{table}[t]
\centering
\caption{Monte Carlo Parameter Dispersions}
\label{tab:mc_dispersions}
\footnotesize
\setlength{\tabcolsep}{1.5pt}
\begin{tabular*}{\columnwidth}{@{\extracolsep{\fill}}l l l l@{}}
\toprule[2pt]
\textbf{Parameter} & \textbf{Nominal} & \textbf{Distribution} & \textbf{Units} \\
\midrule[1pt]
Payload inertia $J_{\mathrm{p}}$        & \num{1.5e-2} & $\N(\mu,\,\SI{10}{\percent})$               & \si{\kilogram\metre\squared} \\
Payload damping $C_{\mathrm{p}}$        & \num{1.0e-4} & $\N(\mu,\,\SI{50}{\percent})$     & \si[per-mode=symbol]{\newton\metre\second\per\radian} \\
Flywheel inertia $J_{\mathrm{mw}}$    & \num{1.3e-4} & $\N(\mu,\,\SI{10}{\percent})$     & \si{\kilogram\metre\squared} \\
Viscous friction $b_{\mathrm{mw}}$    & \num{2.1e-5} & $\U(1.05,\,3.15)\times10^{-5}$ & \si[per-mode=symbol]{\newton\metre\second\per\radian} \\
Viscous friction $b_{\mathrm{pm}}$    & \num{5.0e-5} & $\U(2.5,\,7.5)\times10^{-5}$   & \si[per-mode=symbol]{\newton\metre\second\per\radian} \\
Coulomb friction $f_{\mathrm{C},\mathrm{pm}}$  & \num{1.0e-2} & $\N(\mu,\,\SI{20}{\percent})$               & \si{\newton\metre} \\
Float altitude $z$           & ---          & $\U(1.0,\,25.0)\times10^{3}$   & \si{\metre} \\
\bottomrule[1pt]
\multicolumn{4}{@{}p{\columnwidth}@{}}{\scriptsize $\mu$: nominal value, normal dispersions give $\sigma$ as a percentage of $\mu$. \newline All distributions are clipped at zero.}
\end{tabular*}
\end{table}

\begin{figure}[t]
    \centering
    \includegraphics[width=\linewidth]{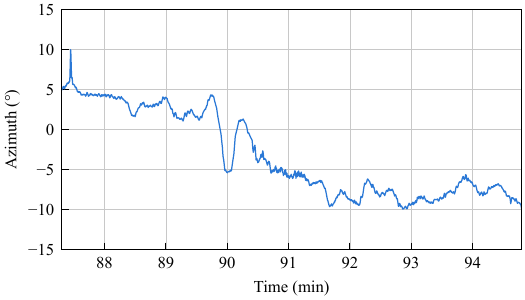}
    \caption{Azimuth error during longest pointing window from May~27 flight data.}
    \label{fig:pointing_window}
\end{figure}

A Monte Carlo campaign is run for each pointing system. The cases with the momentum wheel operated alone and with a pivot motor were compared for pointing performance, momentum management and experiment requirements. The payload is simulated floating for \SI{10}{\minute} at various altitudes to capture performance across the ascent. Results from the simulation can be seen in Fig.~\ref{fig:mc_windows}.

With the pivot motor, 46 runs out of 50 remain within the pointing bounds for more than \SI{97.5}{\percent} of the simulation duration, compared to none without it. The momentum wheel on its own cannot efficiently manage momentum for \SI{600}{\second} and will inevitably saturate, disrupting pointing. This behavior can be observed in Fig.~\ref{fig:mc_windows}(b) where the same number of runs with the pivot motor experience no wheel saturation while essentially all simulations with the wheel-only system experience multiple instances of saturation. The impact on performance can be inferred from Fig.~\ref{fig:mc_windows}(c) where fewer pointing windows are now available for reliable imaging. 

\subsection{Flight Results}
\subsubsection{Flight 1} 
The first flight was on May~7, 2026. The pivot motor was not included on this flight, instead a double-bearing connection kept the flight train and payload decoupled. The balloon reached an altitude of \SI{22.75}{\kilo\metre} after 1 h 10 min in flight. The objective of this flight was to measure the wind disturbances on the payload as angular accelerations, their magnitude and frequency profile. This dataset was used to train disturbance generation models. The flight also measured the payload's tilt stability, which is the root-mean-square (RMS) value of pitch and roll about the payload's nadir axis. An RMS of \ang{12.3} is measured.

\subsubsection{Flight 2}
The first test flight of the momentum wheel was on May~27, 2026. The balloon reached a maximum altitude of \SI{23.5}{\kilo\metre} after an hour and 35 minutes in flight. The pointing system flown consisted of the momentum wheel only, using control modes as described in Fig.~\ref{fig:pointing_fsm} to restore pointing and desaturate the momentum wheel in the absence of a pivot motor. The payload was pointed at north or \ang{0} in the Pixhawk's north-east-down frame. The collected data informed tuning of the pivot motor controller gains and loop rate at different altitudes. The objective of this flight was to measure how long it would take the momentum wheel to saturate at different altitudes, controller performance and tilt stability. 

The flight proved successful: the momentum wheel maintained stable azimuth pointing for extended durations and proved it could desaturate itself and restore pointing reliably. The momentum wheel on its own achieved the experiment's pointing requirements but the benefit of a pivot motor to ensure continuous and uninterrupted pointing is still desired. Looking at Fig.~\ref{fig:pointing}, the disruptions in pointing are clear and they decrease with altitude. After the first 50 minutes or at an altitude of around \SI{11.5}{\kilo\metre} is where the first extended pointing window of 7 minutes and 27 seconds is identified. The final window before the balloon burst was the longest stable window, at 7 minutes and 55 seconds. The payload azimuth angle during this window is plotted in Fig.~\ref{fig:pointing_window}, and a pointing RMSE of \ang{5} is calculated. The cause of the payload azimuth drift visible in this figure is identified in Section~\ref{sec:control} as a limitation of the control strategy, which is intended to be remedied with the addition of a pivot motor.
Throughout the entire flight, 53 windows of \SI{30}{\second} duration pointing within $\pm\SI{8}{\degree}$ of the target azimuth were identified. The pointing RMS error over all the windows is \ang{3.7}. The tilt RMS over the entire ascent is \ang{10.1}, an improvement over the first flight.

\section{Conclusion and Future Work}
\label{sec:lessons}
A system for pointing a lightweight high-altitude balloon payload at a ground-based telescope has been developed and successfully flown. A high-fidelity simulation environment including models for the payload's attitude dynamics, actuators, sensors, and disturbances has been developed and its results reported. A momentum wheel has been successfully flown proving full azimuth control, stabilization and motor desaturation can be accomplished with a single motor. Simulation is validated against flight data to demonstrate fidelity. A desaturation mechanism has been built and a control law has been developed with simulation results to prove it can drastically improve pointing performance. 

The next flight will integrate the pivot motor into the pointing system to demonstrate momentum management in flight. Higher dimensional simulation that accounts for the payload's pendulating motion and its coupling to azimuthal motion should be developed along with higher fidelity wind disturbance models. System parameter uncertainty should be reduced through more precise experimentation. Finally, the flight-proven platform enables iterative testing of alternative non-linear or Lyapunov-based control strategies, to improve pointing accuracy and find more optimal control strategies.

\bibliographystyle{IEEEtran}
\bibliography{references}

\thebiography

\begin{biographywithpic}
{Yorgo Abou Haidar}{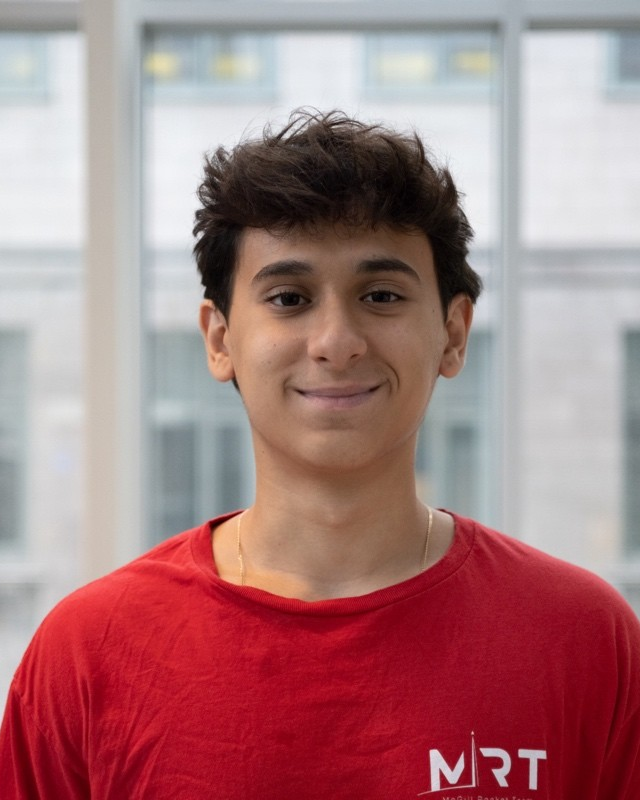}
is an undergraduate student in the McGill Interstellar Flight Experimental Research Group at McGill University. He is completing the B.Eng in Mechanical Engineering. His interests lie in dynamics, estimation and control of spacecraft, orbit determination, space situational awareness and space domain awareness, and interstellar propulsion technologies. 
\end{biographywithpic}
\vspace{1.5\baselineskip}
\begin{biographywithpic}
{James Forbes}{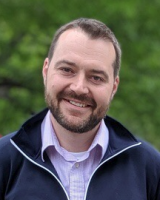}
received the B.A.Sc. degree in Mechanical Engineering (Honours, Co-op) from the University of Waterloo, Waterloo, ON, in 2006, and the M.A.Sc. and Ph.D. degrees in Aerospace Science and Engineering from the University of Toronto Institute for Aerospace Studies, Toronto, ON, in 2008 and 2011, respectively.
He is a Professor in the Department of Mechanical Engineering at McGill University, a Member of the Centre for Intelligent Machines, a Member of the Group for Research in Decision Analysis, a Member of the Centre de Recherches Mathématiques, a Member of the Trottier Institute for Sustainability in Engineering and Design, and an Associate Academic Member of Mila. He holds a William Dawson Scholar award, and received the McGill Association of Mechanical Engineers Professor of the Year Award in 2016, the Engineering Class of 1944 Outstanding Teaching Award in 2018, the Carrie M. Derick Award for Graduate Supervision and Teaching in 2020, the Samuel and Ida Fromson Outstanding Teaching Award in Engineering in 2024, and the inaugural winner of the Donald L. Mordell EUS Teaching Excellence Award in 2026. He is currently a Senior Editor for the International Journal of Robotics Research (IJRR).\end{biographywithpic}
\vspace{1.5\baselineskip}
\begin{biographywithpic}
{Andrew Higgins}{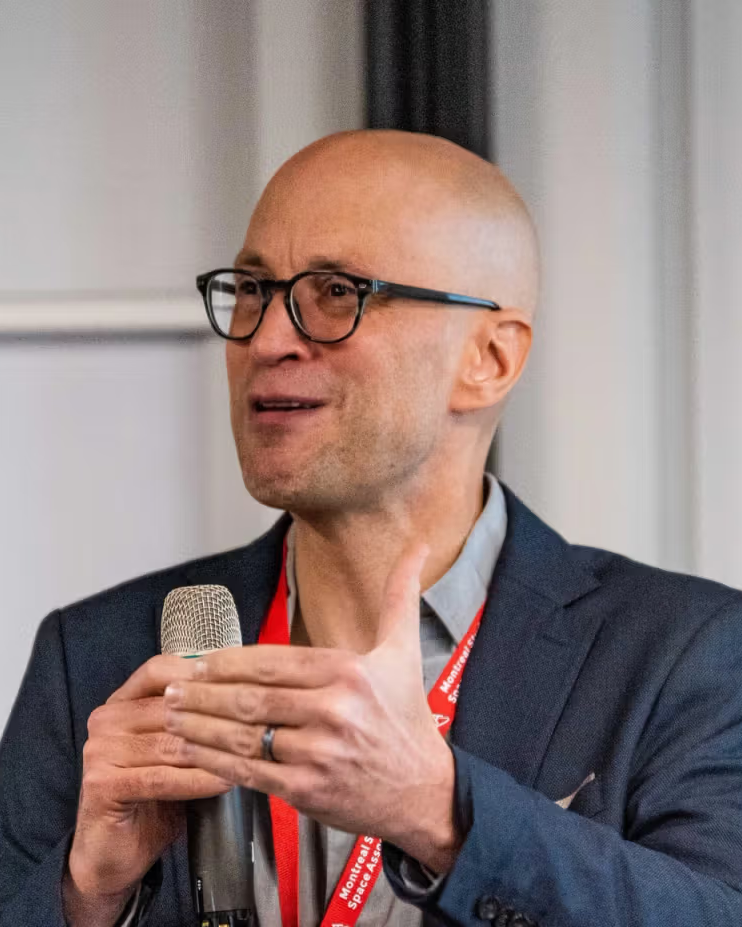}
is a professor of Mechanical Engineering, with Masters and PhD degrees in Aeronautics
and Astronautics from the University of Washington, Seattle. At McGill University, he has taught courses
on compressible flow, thermodynamics, combustion, space propulsion, and statistical mechanics. He has
worked for 30 years on research in high-speed flow, shock waves, and detonation waves for applications
to aerospace propulsion and fusion energy and is now setting his sights higher by considering the
engineering problems associated with interstellar flight.
\end{biographywithpic}
\vspace{1.5\baselineskip}
\begin{biographywithpic}
{Benedetto D'Agnillo}{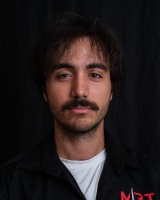}
is an MSc student in Mechanical Engineering part of the Interstellar Flight Experimental Research group at McGill University. He received the BSc in physics from McGill University while conducting research related to project ALTAIR’s radiometric calibration and primary avionics. He is now working on high velocity impact diagnostics in collaboration with General Fusion in Richmond, B.C.\end{biographywithpic}
\vspace{1.5\baselineskip}
\begin{biographywithpic}
{Justin Albert}{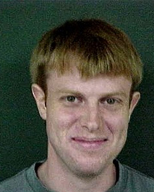}
is an experimental particle astrophysicist and faculty member at the University of Victoria.
His background is in multiple areas of both astrophysics and particle physics, including measurements
of dark matter and dark energy (and precise calibration of the various foregrounds and instrumental
responses that are associated with those measurements); physics of heavy flavor decays \& CP violation;
hadron collider physics; and neutrino physics. Justin is director/PI for the ALTAIR high-altitude micro-airship project for precision photometry calibration for dark energy measurements; and co-PI \& lead
physics \& astronomy faculty member for the ORCASat CubeSat mission, also for precision photometry
calibration; both of which are funded by the Canadian Space Agency and NSERC.\end{biographywithpic}

\end{document}